\documentstyle[12pt]{article}

\begin{document}
\begin{titlepage}
\begin{center}
\today     \hfill    HUTP-95/A040 \\
          \hfill    hep-th/xxxxx \\

\vskip .5in

{\large \bf One-loop Correction and the Dilaton
Runaway Problem } \\
{\bf Rulin Xiu} \\

{\em Lyman Laboratory Of Physics\\
    Harvard University\\
Cambridge, MA 02138, USA}

\end{center}

\vskip .5in

\begin{abstract}
We examine the one-loop vacuum structure of an effective theory of
gaugino condensation coupled to the dilaton for string models
in which the gauge coupling constant does not receive string
threshold corrections.
The new ingredients in our treatment are that
we take into account the  one-loop correction to the dilaton
K\"ahler potential and we use a formulation
which includes a chiral field $H$ corresponding to the gaugino bilinear.
We find through explicit calculation that
supersymmetry in the Yang-Mills sector is  broken by
gaugino condensation.
 The dilaton and $H$ field have masses on the order
of the gaugino condensation scale independently of the dilaton VEV.
Although the calculation performed here is at best a model of
the full gaugino condensation dynamics, the result shows that
the one-loop correction to the dilaton K\"ahler potential as well as
the detailed dynamics at the gaugino condensation scale  may
play an important role in solving the dilaton runaway problem.
\end{abstract}
\end{titlepage}

\setcounter{section}{0}
\section{Introduction}

The determination of the dilaton ($S$) and moduli VEVs is an important
problem in  string phenomenology because it is
directly related to the predictions
of the string models \cite{wdr}.
The dilaton and moduli are flat directions to all orders
in string perturbation theory, and it is hoped that some non-perturbative
effects will lift these flat directions and determine these VEVs dynamically.
Non-perturbative gaugino condensation seems to be a ready candidate for doing
this, but so far much effort has failed to achieve convincing success.
A major reason is that the determination of dilaton VEV through gaugino
condensation suffers from the dilaton runaway problem \cite{wit2},
{\em i.e.} the potential energy is minimized at either $S \rightarrow 0$ or
$S \rightarrow \infty$.
There are several proposals to solve the problem at tree-level in
the observable sector fields.
One is c-number solution \cite{wit2,subr2,rulin},
another is multiple gaugino condensation \cite{kay,dixon,tom3,cas}.
In ref.~\cite{mary2} it is pointed out that another necessary condition for
preventing the dilaton from running away at tree-level is
that the potential energy is positive semi-definite.
This is because the K\"ahler potential for the dilaton is
usually assumed to be $K(S,\bar{S}) = -\log(S + \bar{S})$ and
the potential energy is  proportional to $1 / (S+\bar{S})$.
If the potential energy is negative, it would always have a minimum at
$S=0$  and the dilaton still runs away.
The no-scale structure, which was proposed to naturally suppress
the cosmological constant \cite{no}, is so far the most natural way to
yield the positive semi-definite potential energy. We therefore
focus on the models with the no-scale structure in this paper.

In no-scale models, the dilaton corresponds to a flat direction
at tree level; one-loop effects will lift this degeneracy.
But for many string models, the dilaton VEV determined by
the one-loop potential energy is rather small.
The dilaton VEV consistent with
the weak scale measurement is $\langle S \rangle \sim 2$, which
corresponds to a (model-dependent) hierarchy of at least an order of
magnitude between the Planck scale and the gaugino condensation.
It appears difficult in the present formulation
to dynamically generate such a hierarchy
since the only scale in the theory is the Planck
scale or the string scale.
Because of this, in many models it is found that the gaugino condensation scale
as determined by the dilaton VEV is again on the Planck scale, i.e.
$\langle e^{-S/2b_0}\rangle \sim 1$.

These features are illustrated in the simplest
c-number  gaugino condensation model
\begin{eqnarray}
K &=& -\log(S + \bar{S}) - 3 \log(T+\bar{T}-|\phi|^2), \\
W &=& c_0 + a e^{-3S / 2 b_0}.
\end{eqnarray}
At tree level, dilaton potential is flat and the one-loop corrections
give a potential which is minimized at
$\langle e^{-S / 2b_0} \rangle \sim 1$.
In this model, although one avoid the dilaton runaway problem at the
tree-level, but it ``runs too little'' at the one-loop level.
In this sense, the dilaton runaway appears to be a generic problem in
many models.

In this work we examine the one-loop structure of the formulation
for the dynamics of gaugino condensation coupled to the dilaton
first proposed in ref.~\cite{mary1}, in which a chiral field $H$
corresponding to the gaugino bilinear is introduced.
The new ingredient of our analysis is that we take into account the
one-loop corrections to the dilaton K\"ahler potential.
The inclusion of these effects leads to conclusions that
are quite different from the usual scenario of gaugino condensation.
We find that the dilaton mass as well as that of the $H$ field are on
the same order as the gaugino condensation scale.
This indicates that the usual approach to determining the dilaton VEV
which treats the dilaton as a light field below the gaugino condensation
scale may be incorrect.
We also find that the supersymmetry in the Yang-Mills sector is
broken by gaugino condensation,
unlike the pure Yang-Mills theory first discussed in \cite{vani}.
These results may shed some lights on the dilaton runaway problem.
The  computations also show that the loop-correction to the dilaton
K\"ahler potential may lead to determining  the
dilaton VEV at a value consistent with weak scale measurements.

This paper is organized as follows:
In section 2, we will derive in more detail the
no-scale formulation of gaugino condensation proposed in
ref.~\cite{mary1}.
We show that the inclusion of the one-loop correction to the
dilaton K\"ahler potential and the introduction of a chiral field
related to the gaugino bilinear are  crucial for making the model of
the no-scale type.
In section 3, we give the necessary formulas for
the analysis of the one-loop effective theory which is developed
in \cite{mary3}.
In section 4 and 5, we will analyze the one-loop vacuum
structure for our model.
We will do this in two different ways.
In section 4, we treat  the gaugino-bilinear chiral fields as
heavy field and we integrate it out by solving its equation
of motion.
In section 5, we treat it as a dynamical field with mass
comparable to the dilaton.
We will find that the second analysis yields a minimum at the finite
dilaton VEV,  giving a new scenario for gaugino condensation.
Section 6 contains our conclusions.

\section{No-Scale Formulation}
In this section, we discuss the no-scale formulation of gaugino condensation
for the string models in which the gauge coupling constant
does not receive string threshold corrections.
We are particularly interested in the
models with the no-scale structure
because they naturally suppress the cosmological constant \cite{no}, prevent
the dilaton from running-away at tree level \cite{mary2}, and also because
they might be a natural symmetry for generating the large mass hierarchy
between $M_{\sl GUT}= 0^{16}$ GeV and $M_{\sl SUSY} = 1$ TeV
\cite{rara,rulin}.
In \cite{rulin} it is shown that in a modified c-number model,
the induced constant term in the superpotential is not quantized;
furthermore this supersymmetry breaking scheme
makes it possible to construct affine level one $SU(5)$ or $SO(10)$ string
models with the intermediate gauge symmetry breaking scale
$M_{\sl GUT} = 10^{16}$ GeV.
We therefore restrict attention to the c-number supersymmetry
breaking schemes.

The modular-invariant formulation of effective action with gaugino
condensation coupled to the dilaton field has been discussed in
\cite{modu1,mivr1,mivr2}
for string models in which the gauge
coupling constant receives string threshold corrections.
The no-scale formulation for this type of string models is given
in \cite{mary2}.
The one-loop analysis of these models are given in \cite{mary3}.
In \cite{mary1}, the no-scale formulation of the supersymmetry
breaking dynamics is given
for the string models in which the gauge coupling constant
does not receive string threshold corrections.
The one-loop analysis of  these models has not been carried out so far;
this is the subject of this work.

In the following, we will give a more detailed derivation of
the no-scale formulation of gaugino condensation for the string models that
do not receive string threshold corrections.
We will show that the one-loop correction to
the dilaton K\"ahler potential and the introduction of the $H$ field
play a crucial role in the no-scale structure,
and in preventing the dilaton from running away at tree level.

It has been shown in ref.~\cite{sfer} that the one-loop contribution
to the gauge kinetic terms should be viewed as a field-dependent
wave-function renormalization of the dilaton field rather than a
renormalization of gauge coupling function $f$.
For example, to cancel the modular anomaly the K\"ahler potential is
modified to be
\begin{eqnarray}
K & = & -\log\left(S+\bar{S}-\frac{2b_0}{3}k\right) - k, \\
k & = & -3\log(T+\bar{T}-|\phi|^2),
\end{eqnarray}
while the gauge coupling function remain unchanged:
\begin{equation}
f = S.
\end{equation}
This is usually called Green--Schwarz mechanism \cite{green,sfer,lop,tom1}.
The above K\"ahler potential is obviously not of the no-scale type.
Now one takes into account the full one-loop contribution to
the dilaton  K\"ahler potential which has the form:
\begin{eqnarray}
K = -\log \left(S+\bar{S}+2b_0\log\Lambda^2-\frac{2b_0}{3}k\right)-k,
\end{eqnarray}
where $\Lambda$ is the renormalization scale.
The gaugino condensation scale corresponds to
$\Lambda^2=e^{k/3}|H|^2$, where $H$ is the chiral field relating to the
gaugino bilinear $\langle \lambda \bar{\lambda} \rangle$
(the factor $e^{k/3}$ is included to make $\Lambda$  modular invariant).
We then obtain
\begin{eqnarray}
K &=& -\log \left(S+\bar{S}+2b_0\log|H|^2\right)-k,.
\end{eqnarray}
This model is of no-scale type provided that the superpotential is
independent of the moduli fields.

The superpotential for the $H$ field can be obtained from symmetry
arguments \cite{vani,local,glo,lo}:
\begin{equation}
W = d\left[\frac{1}{4}SH^3+\frac{b_0}{2}H^3\log(\eta H)\right]+c_0+W_0.
\end{equation}
Here $c_0$ and $W_0$ are the contribution from the charged background VEVs
and matter fields.
The parameters $d$ and $\eta$ are not fixed by symmetry requirements,
and specified by the underlying gaugino condensation dynamics.
Under modular transformations,
\begin{eqnarray}
T &\mapsto& T' = \frac{aT-ib}{icT+d},
\\
\Phi^i  &\mapsto&  \Phi^{i\prime} = \frac{\Phi^i}{icT+d},
\\
S &\mapsto& S' = S + 2b_0(icT+d),
\\
H &\mapsto&  H'=\frac{H}{icT+d},
\\
c_0 &\mapsto& c_0'= \frac{c_0}{(icT+d)^3}, \qquad ad-bc=1,
\end{eqnarray}
so that
\begin{eqnarray}
K & \mapsto & K' = K + F + \bar{F}, \\
W & \mapsto & W' = We^{-F}, \\
F & = & 3\log(icT+d).
\end{eqnarray}
For the supergravity theory, the lagrangian depends
on the combination of the K\"ahler potential and superpotential
$G=K+\log|W|^2$,
the above theory is invariant under the modular transformation and
also has the desired no-scale structure.

Although the modification of the dilaton kinetic energy by  loop
effects has been  known for several years,
its consequences for gaugino condensation have not been fully explored.
We see from the above derivation that the inclusion
of the one-loop effects to the dilaton K\"ahler potential is crucial for
the no-scale formulation of the gaugino condensation.
In our following analysis, we find that it has nontrivial effects
on the one-loop vacuum structure and the dynamical determination of
dilaton VEV.

\section{Formalism}
In this section, we write down the necessary formulas for our analysis.
Our calculation largely follows ref.~\cite{mary3},
although (as we explain below) some aspects of the analysis are
quite different.

At tree level, the potential energy can be written as \cite{mary3}
\begin{equation}
V=e^K\left(K_{a\bar{b}}^{-1}\tilde{W}^{a}
\bar{\tilde{W}}^{\bar{b}}\right),
\end{equation}
where $z_{a} = (S, \phi,H)$ and
\begin{equation}
\tilde{W}_{a} \equiv \frac{\partial W}{\partial z_{a}} +
K_{a} W - 3W \frac{K_{a \bar{T}}}{K_{\bar{T}}},
\end{equation}
which manifestly has the no-scale structure.
The tree-level vacuum conditions are
\begin{equation}
\langle \tilde{W}_{a} \rangle =0.
\end{equation}

To calculate the one-loop effective potential, one must first
calculate the mass matrices of the chiral fields.
The scalar squared mass matrix is given by
\begin{eqnarray}
M_S^2&=&\pmatrix{v_{a\bar{b}} & v_{ac} \cr v_{\bar{d}\bar{b}} &
v_{\bar{d}b} \cr}, \\
v_{a\bar{b}}&\equiv&\frac{\partial^2V}{\partial z_a \partial \bar{z}_b}, \\
v_{ab}&\equiv&\frac{\partial^2V}{\partial z_a \partial z_b}-
G_{cd\bar{e}}(G^{-1})^{\bar{e}f}V_f.
\end{eqnarray}
The normalized scalar masses are
\begin{eqnarray}
M_S^{r2}\equiv\pmatrix{G^{-1/2} & 0 \cr 0 & (G^{-1/2})^T \cr}
\pmatrix{v_{a\bar{b}} & v_{ac} \cr v_{\bar{d}\bar{b}} & v_{\bar{d}b} \cr}
\pmatrix{G^{-1/2} & 0 \cr 0 & (G^{-1/2})^T \cr},
\end{eqnarray}
which has the same eigenvalues as
\begin{eqnarray}
\tilde{M}_s^{r2}=\pmatrix{v_{a\bar{b}} & v_{ac} \cr v_{\bar{d}\bar{b}}
& v_{\bar{d}b}\cr}
\pmatrix{G^{-1} & 0 \cr 0 & (G^{-1})^T \cr}.
\end{eqnarray}
Under the tree-level vacuum condition above, one obtains
\begin{eqnarray}
v_{a\bar{b}}&=&e^K \left[
\tilde{W}_{ac}(G^{-1})^{c\bar{d}}\bar{\tilde{W}}_{\bar{d}b}+
\bar{\tilde{W}}_{a\bar{c}}(G^{-1})^{\bar{c}d}\tilde{W}_{d\bar{b}}\right],
\\
v_{ab}&=&e^K\left[
\tilde{W}_{ac}(G^{-1})^{c\bar{d}}\bar{\tilde{W}}_{\bar{d}b}+
\bar{\tilde{W}}_{a\bar{c}}(G^{-1})^{\bar{c}d}\tilde{W}_{db} \right].
\end{eqnarray}

The gaugino mass parameter is given by
\begin{equation}
(M_{1/2})_{\alpha\beta}=\frac{1}{2}e^{G/2}(G^{-1})^{a\bar{b}}G_{\bar{b}}
f_{\alpha\beta,a},
\end{equation}
while the normalized gaugino mass-squared is
\begin{equation}
(M_{1/2}M^{+}_{1/2})_{\alpha\beta}\equiv (M_{1/2})_{\alpha\gamma}
[(Re f)^{-1}]^{\gamma\delta}(M^{+}_{1/2})_{\delta\beta}.
\end{equation}
The fermion masses are given by
\begin{eqnarray}
\mu_{ab}=e^{G/2}\left[G_{ab}+G_aG_b-
\frac{1}{3}G_c(G^{-1})^{c\bar{d}}G_{\bar{d}ab}\right]
\end{eqnarray}
The normalized fermion mass matrix is,
\begin{eqnarray}
m_{ab}=(G^{-\frac{1}{2}})^c_a\mu_{cd}(G^{-\frac{1}{2}})^d_b,
\end{eqnarray}
which has the same eigenvalue as
\begin{equation}
\tilde{m}_{ab}\equiv m_{ac}(G^{-1})^c_b.
\end{equation}
In the above,
\begin{eqnarray}
G_a&=&K_a+\frac{W_a}{W}, \\
G_{ab}&=&K_{ab}+\frac{W_{ab}}{W}-\frac{W_aW_b}{W^2}, \\
G_{a\bar{b}}&=&K_{a\bar{b}}, \\
(G^{-1})^{a\bar{c}}G_{\bar{c}b}&=&\delta^a_b.
\end{eqnarray}

The one-loop potential is
\begin{equation}
V^{1-loop} = \frac{1}{4(4\pi)^2}[2 {\rm Str}(M^2\Lambda^2)+
{\rm Str}(M^4\log(M^2/\Lambda^2))],
\end{equation}
where Str is the supertrace.
In our one-loop analysis, we use
the tree-level vacuum conditions to calculate the one-loop potential
energy, with which we determine
the rest of the VEVs of the scalar fields. This approximation corresponds
to determining the VEVs to  order $\hbar$

\section{Gaugino Bilinear as a Heavy Field}
In this section, we will carry out the one-loop analysis for the
models formulated in Section 2.
Here we treat the gaugino bilinear as a heavy field
(as is usually assumed), {\em i.e.}
 we integrate out the $H$ field
using its equation of motion.

{}From the tree-level vacuum condition, we get the classical
equation of motion for $H$ field,
\begin{equation}
\tilde{W}_H \equiv W_H - \frac{2b_0}{H(S+\bar{S}+2b_0\log|H|^2)} W =0.
\end{equation}
To solve this equation, we write
\begin{equation}
H=h(S) e^{S/2b_0}.
\end{equation}
The function $h$ is determined by
\begin{equation}
d h^3[(3L-2b_0)\log(h\eta)+L]=4c_0e^{3S/2b_0},
\end{equation}
where
\begin{equation}
L=S+\bar{S}+2b_0\log|H|^2=2b_0\log|h|^2.
\end{equation}
It is easy to see that gaugino condensation occurs ($h \ne 0$)
if and only if the constant part of the superpotential $c_0$ is
nonzero.
This is consistent with our assumption that the $c_0$ is induced by
gaugino condensation.
After solving for $H$, we obtain the effective theory below the gaugino
condensation scale:
\begin{eqnarray}
K&=&-\log L+k,
\qquad
k=-3 \log\left[T+\bar{T}-|h|^2e^{(S+\bar{S})/2b_0}\right], \\
W&=&d \frac{b_0}{2} e^{-3S/2b_0}h^3 \log\eta h + c_0 + W_0.
\end{eqnarray}
The tree level potential energy is
\begin{equation}
V=e^K \left(|\tilde{W}_s|^2+|W_i|^2\right),
\end{equation}
which is minimized at
\begin{eqnarray}
\label{vacc}
\langle W_i \rangle &=&0,
\\
\label{vaccc}
\langle \tilde{W}_S \rangle &\equiv&
\langle W_S+K_SW-\frac{3K_{S\bar{T}}}{K_{\bar{T}}}W \rangle
=\langle W_S-\frac{L_S}{L}W \rangle=0.
\end{eqnarray}
Note that in the tree-level potential energy, the $A$ term and scalar masses
of the matter sector remain zero although local supersymmetry is
broken by $\langle W \rangle \ne 0$.
We find that at the tree-level minimum
\begin{equation}
G^S = (G^{-1})^{S\bar{a}}G_{\bar{a}}=0, \qquad
G^T = (G^{-1})^{T\bar{a}}G_{\bar{a}}= -e^{k/3},
\end{equation}
so the gauginos in the matter sector remain massless if the
$f$ function does not depend on the moduli;
these are the models in which the gauge coupling constant
does not receive moduli-dependent string threshold corrections.
We see that at tree-level, global supersymmetry is broken in the
dilaton   sector but not in the matter sector for the
models we are considering.
In this approximation, all the moduli masses vanish.

We now solve the tree-level vacuum conditions eqs.~(\ref{vacc})
and (\ref{vaccc}).
Eq.~(\ref{vacc}) is automatically satisfied if the VEVs
of the matter fields are zero for the tri-linear superpotential.
Eq.~(\ref{vaccc}) imposes one relation between $c_0$ and $S$;
the dilaton and moduli fields correspond to flat directions
which might be lifted by loop effects.
Solving eq.~(\ref{vaccc}), we obtain
\begin{equation}
\langle h \rangle=\eta^{-1}.
\end{equation}
The parameter $\eta$ is determined by the gaugino condensation
dynamics and is expected to be of order 1.
Now we can also determine $c_0$ in terms of $S$,
\begin{equation}
\langle W \rangle = c_0 = \frac{1}{4} \langle d L h^3 e^{-3S/2b_0} \rangle.
\end{equation}

We obtain the dilaton scalar squared masses
\begin{eqnarray}
 M_S^2 &=& m_{3/2}^2 \left[
(G^{-1})^{S\bar{S}}\frac{1\pm 6L/(2b_0)}{4L^2}
\right]^2
\nonumber\\
&=& \frac{(1\pm 6L/(2b_0))^2}{(1+3xL^2/(4b_0^2))^4},
\end{eqnarray}
and the dilaton fermion mass
\begin{eqnarray}
 M_f &=& m_{3/2} (G^{-1})^{S\bar{S}}\frac{3}{4 b_0 L} \\
     &=& m_{3/2} \frac{3L/b_0}{(1+3L^2x/4b_0^2)^2},
\end{eqnarray}
with
\begin{eqnarray}
(G^{-1})^{S\bar{S}}&=&\frac{4L^2}{(1+
3L^2 x / 4b_0^2)^2}, \\
m^2_{3/2}&=&e^K|W|^2=\frac{1}{16}d^2Lx^3,\\
x &=& \frac{|h|^2}{k}e^{-(S+\bar{S})/2b_0}.
\end{eqnarray}
The cutoff is
\begin{eqnarray}
\Lambda^2 &=& M_S^2|H|^2e^{K/3}
\nonumber\\
&=&\frac{2}{S+\bar{S}+2b_0\log(T+\bar{T})}|H|^2e^{K/3}
\nonumber\\
&=&\frac{2x}{L+2b_0 \log(x+1)},
\end{eqnarray}
where
\begin{equation}
M_S^2=\frac{2M^2_p}{S+\bar{S}+2b_0\log(T+\bar{T})}
\end{equation}
is the string scale, which is also
the scale at which the string gauge coupling constants
``unify'' \cite{kap,mary4}.
We can now obtain the one-loop potential energy, which depends only
on the moduli-invariant dilaton and moduli combination
$x$:
\begin{eqnarray}
V&=&2[-4+\frac{2}{(1+3L2x/4b_0^2)^2}]\frac{Ld^2}{16}x^4  \\ \nonumber
& + &(\frac{Ld^2}{16})^2x^6
\{[-4+\frac{2+12(6L/2b_0)^2}{(1+3L^2x)^4}]\log \left[
\frac{d^2L}{32}(L+2b_0 \log(x+1))x^2\right] + g\} \\
g &=& \frac{1}{(1+3L^2x/4b_0^2)^4}[(1+6L/2b_0)^4\log(1+6L/2b_0)^2
\\ \nonumber
&+&(1-6L/2b_0)^4\log(1-6L/2b_0)^2-2(6L/2b_0)^4\log(6L/2b_0)^2].
\end{eqnarray}
We find that the minimum of  the above one-loop potential energy
is reached at either $x\rightarrow \infty$ or $x=0$ depending the different
ways of minimizing the potential. In any case,
the model suffers the dilaton runaway problem.

In the next section, we consider the possibility that the dilaton mass is
on the same order as the gaugino condensation scale, and find by an
explicit calculation that this is a real possibility.

\section {Gaugino Bilinear As A Dynamical Field}
In this Section, we study the one-loop vacuum structure of our model
treating the $H$ field as a dynamical field.
As already pointed out, this model cannot be viewed as anything more
than a model of the dynamics at the gaugino condensation scale, similar
in spirit to the linear sigma model as a model of QCD.
Still, it is a reasonable model, and the qualitative conclusions may
be correct.

In this case, the dilaton field and the $H$ field mix, and the
inverse of the K\"ahler metric is
\begin{equation}
(G^{-1})^{i\bar{j}}=\pmatrix{L^2 + 4b^2C / (3|H|^2) &
         -2bC / (3H) & -\frac{2}{3} bC \cr
         -2bC/{3\bar{H}} & \frac{1}{3} {C} & \frac{1}{3} {HC} \cr
         -\frac{2}{3} {bC}& \frac{1}{3}{\bar{H}C}& \frac{1}{3}C(|H|^2+C) \cr},
\end{equation}
where
\begin{equation}
L=S+\bar{S}+2b_0\log|H|^2,
\qquad C=T+\bar{T}-|H|^2.
\end{equation}
Solving the tree-level vacuum condition
\begin{eqnarray}
\label{tvacc}
\langle \tilde{W}_H \rangle \equiv \langle W_H-\frac{L_H}{L}W\rangle=0,
\\
\label{tvaccc}
\langle \tilde{W}_S \rangle \equiv\langle W_S+K_SW\rangle =0,
\end{eqnarray}
we get
\begin{eqnarray}
H &=& \frac{1}{\eta} e^{-s / (2b_0)}
= h e^{-s /(2b_0)},
\\
\langle W \rangle &=& c_0 = \frac{1}{4} \langle dLH^3 \rangle.
\end{eqnarray}
The orders of magnitude of the solutions are
\begin{eqnarray}
H &\sim& e^{-s / (2b_0)}, \\
\langle W \rangle &\sim& e^{-3s / (2b_0)},
\end{eqnarray}
so  $h\sim 1$, $d_0\equiv \frac{1}{4}dL=\frac{1}{4}d\log(2b_0|h|^2) \sim 1$.
We take the point of view that $h$ and $d_0$ are constants to be
determined from a better understanding of the gaugino condensation
dynamics.
This is different from the point of view of ref.~\cite{mary3},
where these parameters are taken to be dynamical variables
to be determined by minimization of the effective potential.

{}From the vacuum conditions eqs.~(\ref{tvacc}) and (\ref{tvaccc}) we
obtain
\begin{eqnarray}
\langle G^i \rangle = \langle G^{i\bar{k}}G_{\bar{k}} \rangle =
\pmatrix{0&0&-C\cr},
\end{eqnarray}
so the tree-level gaugino mass is zero in the models we are
discussing.
With the above relations, we obtain the fermion mass matrix
\begin{eqnarray}
(\mu_{1/2})_{IJ}=e^{{G}/{2}}
\pmatrix{0 & {3}/(LH)&0\cr
             {3}/(LH)& {12b_0} / ({H^2L}) &0\cr 0&0&0\cr}.
\end{eqnarray}
The normalized fermion mass matrix is
\begin{eqnarray}
&& (m_{1/2})_{IJ} = (\mu_{1/2})_{IJ}(G^{-1})^{J\bar{K}}
\nonumber\\
&=& e^{{G}/{2}}
\pmatrix{-{2bC}/({L|H|^2})&{c}/({LH})&{C}/{L}\cr
{3L}/{H} - {4b_0^2C}/ ({|H|^2HL}) & {2b_0C}/({H^2L})&
{2b_0C}/({HL})\cr 0&0&0\cr}.
\end{eqnarray}
Assuming that $S=\bar{S},H=\bar{H}$ ({\em i.e.}\ $CP$ violation is
highly suppressed) we obtain the fermion mass eigenvalues
\begin{eqnarray}
m_{1}^{1/2}=m_{2}^{1/2}=e^{{G}/{2}}\frac{\sqrt{3C}}{HL},
\qquad m^{1/2}_3=0.
\end{eqnarray}
The scalar masses are computed from
\begin{eqnarray}
\tilde{W}_{i\bar{k}}&=&\pmatrix{1&{2b_0}/({\bar{H}})&0\cr
{2b_0}/{H}&{4b_0^2}/{|H|^2}&0\cr 0&0&0 \cr} {W}/{L^2},
\\
\tilde{W}_{ik}&=&\frac{3}{4}H^2
\pmatrix{0&1&0\cr 1&{4b_0}/{H}&0 \cr
0&0&0 \cr},
\\
v_{a\bar{b}}&=&e^G\pmatrix{\frac{3C+|H|^2}{|H|^2L^2} &
\frac{2b_0(3C+\bar{H}^2)}{|H|^2\bar{H}L^2}&0\cr
\frac{2b_0(3C+H^2)}{|H|^2HL^2}&\frac{4b_0^2(3C+|H|^2)+
9|H|^2L^2}{|H|^4L^2}&0\cr 0&0&0\cr}, \\
v_{ab}&=&e^G\pmatrix{0&\frac{3}{HL}&0\cr \frac{3}{HL}&
\frac{12b_0}{H^2L}&0\cr 0&0&0\cr}.
\end{eqnarray}
Again assuming $S=\bar{S}$, $H=\bar{H}$, the normalized mass matrix
has the same eigenvalues as
\begin{equation}
\tilde{M}_s^{r2}\equiv\frac{1}{2}e^G\pmatrix{1&-1\cr 1&1}
\pmatrix{v_{a\bar{b}}G^{-1}&v_{ac}G^{-1}\cr
v_{\bar{d}\bar{b}}G^{-1}&v_{\bar{d}b}G^{-1}\cr}
\pmatrix{1&1\cr -1&1}.
\end{equation}
With this trick, the scalar masses can be easily obtained:
\begin{eqnarray}
M^{s2}_{1,3}&=&e^G\frac{|H|^2(6C+|H|^2)+H^3\sqrt{12C+|H|^2}}{2|H|^4}, \\
M^{s2}_{2,4}&=&e^G\frac{|H|^2(6C+|H|^2)-H^3\sqrt{12C+|H|^2}}{2|H|^4}, \\
M^{s2}_{5,6}&=& 0.
\end{eqnarray}
Here the gravitino mass is
\begin{equation}
m_{3/2}^2=e^{G}=e^K|W|^2=\frac{d_0^2 |H|^6}{L C^3}.
\end{equation}
The moduli scalar masses are also zero as expected.

The above result indicates that the inclusion of one-loop correction
to the dilaton K\"ahler potential yields several interesting features.
The dilaton mass is equal to the $H$ mass (to be identified
with the scale of gaugino condensation) {\em independently of the
value of the dilaton VEV}.
Also, supersymmetry is broken in the hidden sector.
We will discuss the possibility of obtaining a hierarchy between the
gaugino condensation scale and the Planck (or string) scale below.

We also calculate the $H$ field and dilaton masses in a more general
class of models which have the same K\"ahler potential but different
superpotential:
\begin{equation}
W = d e^{-3S/2b_0}Y^n \log(\eta Y), \qquad
Y = e^{S/2b_0} H.
\end{equation}
(The case $n=3$ corresponds to the model discussed above.)
In this class of models, we find the fermion masses are
\begin{eqnarray}
m^f = e^{G/2}\left\{
\frac{(n-3)L}{2b_0}\pm \sqrt{\left[\frac{(n-3)L}{2b_0}\right]^2+3z}
\right\}
\end{eqnarray}
and the scalar masses are
\begin{eqnarray}
M^{s2}_{1,3} &=& e^G \Biggl\{
3z+\frac{1}{2}\left[1+(3-n)\frac{L}{b_0}\right]^2
\nonumber\\
&& \quad \pm \frac{1}{2}
\left[1+(3-n)\frac{L}{b_0}\right]
\sqrt{12z+\left[1+(3-n)\frac{L}{b_0}\right]^2} \,\Biggr\},
\\
M^{s2}_{2,4} &=&
e^G \Biggl\{
3z+\frac{1}{2}\left[1-(3-n)\frac{L}{b_0}\right]^2
\nonumber\\
&& \quad \pm \frac{1}{2}
\left[1-(3-n)\frac{L}{b_0}\right]
\sqrt{12z+\left[1-(3-n)\frac{L}{b_0}\right]^2} \,\Biggr\},
\\
M^{s2}_{5,6} &=& 0,
\end{eqnarray}
here $z={C}/{|H|^2}$. For $n\neq 3$,
the dilaton and the $H$ field masses are no longer equal,
but they are still the same order of magnitude for any value of the
dilaton VEV.
Futhermore, supersymmetry is still broken in the Yang-Mills sector.

With the above mass matrixes calculated, we can easily write down
the one-loop vacuum potential energy.
The potential energy only depends on the modular
invariant function $z={C}/{|H|^2}$ so the moduli and dilaton VEVs are
not uniquely determined in this model.
We take the cut-off scale to be
\begin{eqnarray}
\Lambda^2 &=& M_S^2|H|^2e^{K/3}
\nonumber\\
&=&\frac{2}{S+\bar{S}+2b_0\log(T+\bar{T})}|H|^2e^{K/3}
\nonumber\\
&=&\frac{2z^{-1}}{L+2b_0 \log(z+1)},
\end{eqnarray}
where
\begin{equation}
M_S^2=\frac{2M^2_p}{S+\bar{S}+2b_0\log(T+\bar{T})}
\end{equation}
is the string scale, which is also
the scale at which the string gauge coupling constants
``unify'' \cite{kap,mary4}.
The one-loop effective potential is
\begin{eqnarray}
v_1 &=& 64\pi^2V^{\rm 1-loop}
\nonumber\\
&=&-\frac{8d_0^2L^{-1}}{L+2b_0 \log(z+1)}z^{-4}
\nonumber\\
&& +\, d_0^4z^{-6}L^{-2}
\left\{ (-2+24z)\log\left[d_0z^{-2}
\left(\frac{1}{2}+b_0/L\log(z+1)\right)\right] + g_0 \right\}
 \nonumber \\
g_0 &=& -4(3z)^2\log(3z)+2
\left(\sqrt{3z+\frac{1}{4}}+\frac{1}{2}\right)^4
\log(\sqrt{3z+\frac{1}{4}}
+ \frac{1}{2})^2 \\ \nonumber
&+&2\left(\sqrt{3z+\frac{1}{4}} -\frac{1}{2}\right)^4
\log\left(\sqrt{3z+\frac{1}{4}}-\frac{1}{2}\right)^2
\end{eqnarray}
The minimum condition of the potential energy  determine $z$ as a
function of $d_0$ and $h$.
The result is that for fixed $d_0 \sim 1$, $z$ is a smooth
function of $h$ which diverges at $h = 1$ as
\begin{equation}
\frac{z}{ \log z} \propto \frac{ b_0 d_0}{L} = \frac{d_0}{4 \log h}.
\end{equation}
We know that $L$ is related to the one-loop
gauge coupling constant $g^2$ at the gaugino
condensation scale through: $2/L = g^2$. One can see that the large
$g^2$ might yield to the large  $z$, thus the hierarchy between the
string scale and the gaugino condensation scale. Since  the gauge
coupling constant at the gaugino condensation scale is very large,
the higher-loop correction  becomes
very important. The conclusive result will depend on  the inclusion of
the higher-loop corrections to the dilaton K\"ahler potential. We will
defer this discussion to the future work.

\section{Conclusion}
We have shown that the one-loop correction to the dilaton K\"ahler potential
may significantly change the dynamics of gaugino condensation coupled to
the dilaton.
In a specific model including a dynamical field $H$ for the gaugino
bilinear, we find that the supersymmetry is broken by gaugino condensation
in the Yang-Mills sector.
We also find that the dilaton and the $H$ field have masses on the same
order of magnitude as the gaugino condensation scale.
We thus propose that the determination of the dilaton
VEV through gaugino condensation may depend sensitively on the dynamics
at the gaugino condensation scale.
This is very different from the usual scenario in which the dilaton is
lighter than the gaugino condensation scale, and is treated as a light
field in an effective theory below the gaugino condensation scale.

We also find that the large value of the gauge coupling
 at the gaugino condensation scale
might lead to a
hierarchy between the string scale and the gaugino condensation
scale, and fix the dilaton VEV at a realistic value.
The higher loop corrections to the dilaton K\"ahler potential
may also be important and we  defer the investigation  to the future work.

The main conclusion of this work is that the inclusion of the loop correction
to the dilaton K\"ahler potential may dramatically change the scenario
of determining the dilaton VEV through the gaugino condensation
and may lead to the solution of the dilaton runaway problem.

\vskip 28pt
\noindent{\bf Acknowledgement}
\vskip 12pt

I would like to thank Mary K. Gaillard, Jim Liu and
Markus Luty for very helpful discussions.

\vskip 28pt

\vfill\eject

\end{document}